\documentclass[conference]{IEEEtran}
\IEEEoverridecommandlockouts
\usepackage{threeparttable}
\usepackage{cite}
\usepackage{multirow}
\usepackage{makecell}
\usepackage{booktabs}
\usepackage{amsmath,amssymb,amsfonts}
\usepackage{algpseudocode}
\usepackage{graphicx}
\usepackage{tikz}
\usepackage{hyperref}
\usepackage{textcomp}
\usepackage{subcaption}
\usepackage{xcolor}
\usepackage{pifont}
\usepackage{array}
\def\BibTeX{{\rm B\kern-.05em{\sc i\kern-.025em b}\kern-.08em
    T\kern-.1667em\lower.7ex\hbox{E}\kern-.125emX}}
\begin{document}
\captionsetup{font={small}}
\title{MERba: Multi-Receptive Field MambaVision for Micro-Expression Recognition}
% \title{Multi-Receptive Field Local-Global Feature Integration for Micro-Expression Recognition}
%with Debiased Inference}
%Asymmetric-Scan
% \author{Anonymous BIBM submission}

\author{
	\IEEEauthorblockN{
            Xinglong Mao$^{1}$,
		Shifeng Liu$^{1}$, 
		Sirui Zhao$^{1*}$, 
		Tong Xu$^{1}$, 
            Hanchao Wang$^{2}$,
            Baozhi Jia$^{2}$,
        and Enhong Chen$^{1*}$} 
	\IEEEauthorblockA{$^1$ State Key Laboratory of Cognitive Intelligence, University of Science and Technology of China, Hefei, China}
        \IEEEauthorblockA{$^2$ Reconova Technologies Co., Ltd., Xiamen, China\\(* Corresponding authors)}
	\IEEEauthorblockA{maoxl@mail.ustc.edu.cn, \{siruit,cheneh\}@ustc.edu.cn}
 \IEEEcompsocitemizethanks{\IEEEcompsocthanksitem This work has been submitted to the IEEE for possible publication. Copyright may be transferred without notice, after which this version may no longer be accessible.}}

\maketitle

\begin{abstract}
Micro-expressions (MEs) are brief, involuntary facial movements that reveal genuine emotions, offering valuable insights for psychological assessment and criminal investigations. Despite significant progress in automatic ME recognition (MER), existing methods still struggle to simultaneously capture localized muscle activations and global facial dependencies, both essential for decoding subtle emotional cues.
To address this challenge, we propose MERba, a hierarchical multi-receptive field architecture specially designed for MER, which incorporates a series of Local-Global Feature Integration stages.
% To address this challenge, we propose MERba, which incorporates a series of Local-Global Feature Integration stages with a hierarchical multi-receptive field mechanism. This design enables a progressive transition from fine-grained motion perception to comprehensive facial expression understanding.
Within each stage, detailed intra-window motion patterns are captured using MERba Local Extractors, which integrate MambaVision Mixers with a tailored asymmetric multi-scanning strategy to enhance local spatial sensitivity. These localized features are then aggregated through lightweight self-attention layers that explicitly model inter-window relationships, enabling effective global context construction.
Furthermore, to mitigate the challenge of high inter-class similarity among negative MEs, we introduce a Dual-Granularity Classification Module that decomposes the recognition task into a coarse-to-fine paradigm. Extensive experiments on three benchmark datasets demonstrate that MERba consistently outperforms existing methods, with ablation studies confirming the effectiveness of each proposed component.

\end{abstract}


\section{Introduction}
Micro-expressions~(MEs) are involuntary facial movements that reveal an individual's genuine emotions and intentions, holding significant value in psychological research~\cite{ben2021video}. Unlike regular facial expressions, MEs are characterized by short duration (typically less than 0.5 seconds~\cite{yan2013fast}), low intensity, and occurrence in localized facial regions. Recent advances in ME recognition~(MER) have shown promise in fields such as criminal investigations and psychological diagnostics~\cite{ekman2009telling,huang2021elderly}. 
With the rapid development of artificial intelligence, automatic MER has gained growing attention in the field of affective computing. To extract effective features from MEs, researchers have developed various sophisticated feature extractors based on Convolutional Neural Networks~(CNNs)~\cite{liong2019shallow,zhao2021two,zhou2022feature} and Vision Transformers~(ViTs)~\cite{wei2022novel,cai2024mfdan,wang2024htnet}. While CNNs excel at capturing local features, they have limitations in modeling global dependencies, whereas ViTs handle global information well, but are computationally expensive and require large amounts of data, often leading to severe overfitting in MER.

Recently, the Mamba architecture~\cite{gu2023mamba}, based on State Space Models~(SSMs)~\cite{kalman1960new}, has demonstrated outstanding performance across various computer vision tasks~\cite{zhu2024vision,liu2024vmamba}, particularly favored for its efficient linear time complexity, strong contextual awareness, and simplified network structure.
Building on this foundation, MambaVision~\cite{hatamizadeh2025mambavision} extends Mamba by integrating Transformer blocks, forming a hybrid structure that aims to balance computational efficiency with long-range spatial dependency modeling. However, directly applying MambaVision to MER tasks is suboptimal, which fails to focus sufficiently on localized facial regions where MEs typically manifest as isolated Action Units (AUs) \cite{ekman1978facial} and neglects fine-grained inter-region dependencies. Indeed, our empirical results (as in Table \ref{tab:result}$-$\ref{tab:result_mmew}) show that the baseline MambaVision model underperforms significantly on MER benchmarks, indicating its limited capacity to capture subtle, region-specific motion dynamics.

To tackle the above challenges, we propose MERba, a hierarchical multi-receptive field architecture specifically designed for MER, incorporating the strengths of Mamba and Transformer in a task-adaptive manner. Specifically, MERba integrates fine-grained local feature extraction with global dependency modeling through multiple ME Local-Global Feature Integration~(LGFI) Stages. In each LGFI stage, instead of scanning the entire image, we define segmented non-overlapping local windows, within which MERba local extractors equipped with MambaVision Mixers (Mixers) operate independently to extract intra-window ME motion details. However, local motion perception remains fragmented and lacks contextual relevance. To address this, we employ global self-attention blocks to aggregate these localized features, explicitly modeling inter-window relationships to establish a holistic global context, thereby obtaining meaningful representations of global emotional movements. Furthermore, we employ progressive downsampling between stages to achieve systematic expansion of the local window receptive field, allowing a hierarchical transition from localized motion perception to comprehensive facial expression understanding.

% It is worth noting that prior SSM-based models~\cite{zhu2024vision,liu2024vmamba,huang2025localmamba,yang2024plainmamba} often adopt symmetric or bi-directional scans to enhance spatial coherence. 
It is worth noting that reasonable scanning strategies are critical for enhancing Mamba’s performance on visual tasks, with existing methods often utilizing bi-directional or symmetric scanning directions to better capture spatial contiguity relationships~\cite{zhu2024vision,liu2024vmamba,huang2025localmamba,yang2024plainmamba}. However, due to the fixed facial structure after alignment and the natural left-right symmetry of human faces, such strategies introduce redundancy and computational overhead. To overcome these limitations, we propose an Asymmetric Multi-Scanning Strategy, which uses four asymmetric scanning directions that retain spatial diversity while minimizing unnecessary duplication.

Additionally, to address the challenge of high semantic similarity and inter-class confusion among negative emotions (e.g., anger and disgust), we introduce a Dual-Granularity Classification Module (DGCM). This module uses two parallel classification heads: a coarse classifier that determines emotional polarity (e.g., positive or negative) and a fine classifier that resolves more nuanced emotional categories. This coarse-to-fine paradigm improves the model’s discriminability and robustness, especially under the imbalanced conditions common in ME datasets.

Conclusively, our contributions are summarized as follows:
\begin{itemize}
%\item We propose MERba, a multi-receptive field architecture for MER, which integrates local extractors based on MambaVision Mixers with global self-attention mechanisms, facilitating fine-grained movement learning and global dependency modeling.
%\item We design an asymmetric multi-scanning strategy to eliminate redundant scanning directions and enhance local spatial perception.
%\item We introduce a selectively activated Dual-Granularity Classification Module that enhances fine-grained emotion recognition via a coarse-to-fine classification paradigm.
\item We propose MERba, a novel multi-receptive field architecture for MER, which introduces hierarchical Local-Global Feature Integration stages to enable a progressive transition from detailed localized motion perception to holistic facial understanding.
\item We design an effective local-to-global feature modeling mechanism, where MambaVision Mixers combined with a specialized asymmetric multi-scanning strategy are used to extract fine-grained motion features within non-overlapping windows, and lightweight self-attention layers are employed to capture global dependencies across localized facial regions.
\item We introduce a selectively activated Dual-Granularity Classification Module that decomposes the recognition task into a coarse-to-fine paradigm, effectively reducing inter-class confusion among negative MEs.
\item Our approach achieves state-of-the-art~(SOTA) performance on multiple benchmark ME datasets, with ablation studies validating the effectiveness of each proposed component.
\end{itemize}

\begin{figure}[tbp]
    \centering
    \includegraphics[width=0.48\textwidth]{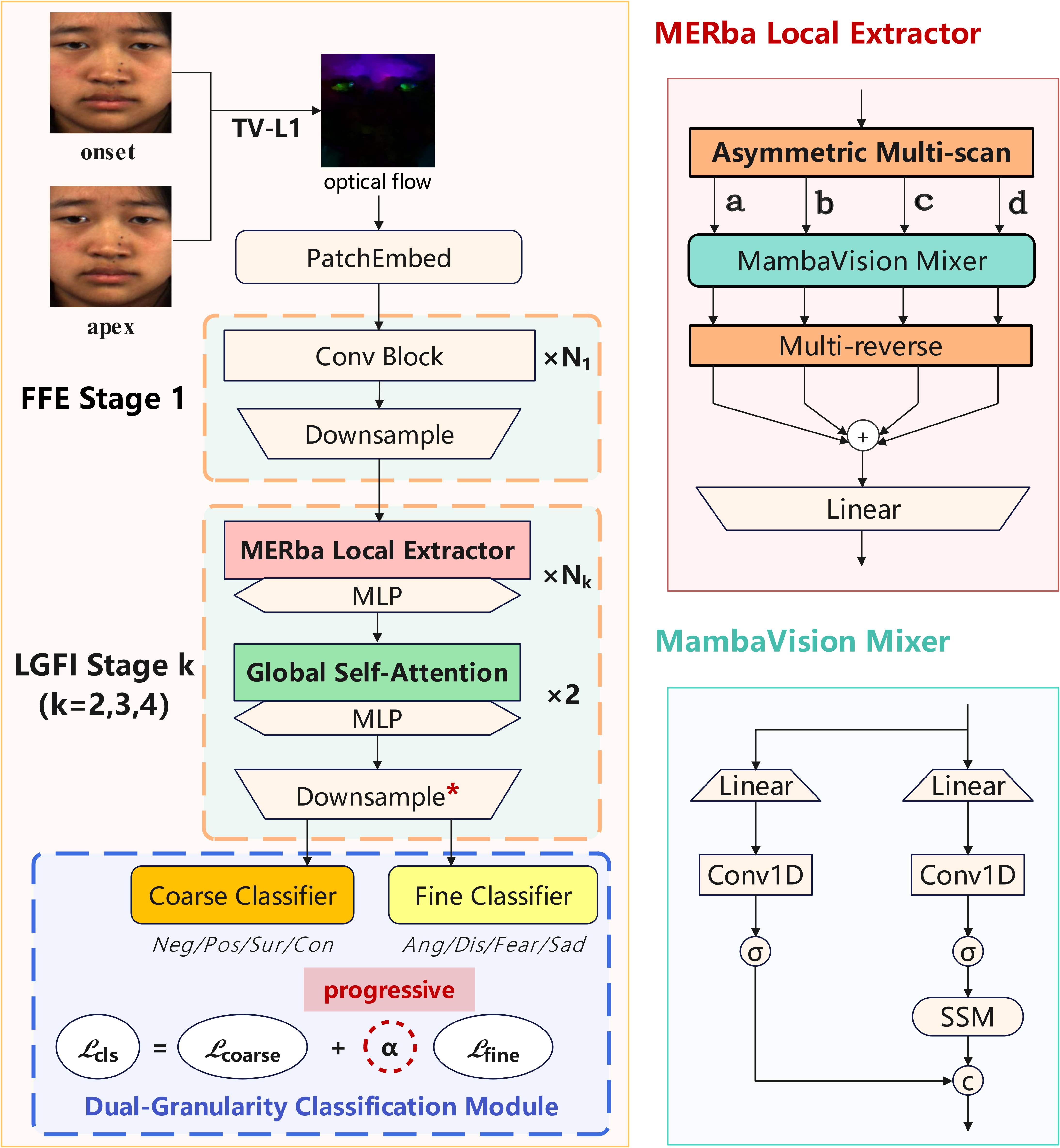}
    \caption{The overall framework of MERba, with the specific structures of the MERba Local Extractor and MambaVision Mixer shown on the right side. * denotes that the Downsample Layer in Stage 4 is replaced with 2D Batchnorm and 2D Average Pooling.}
    \label{fig:pipeline}
\end{figure}
%\vspace{-0.3cm}

\section{Related Work}
\subsection{Automatic Micro-Expression Recognition}
%传统特征到深度特征
Early automatic MER methods primarily relied on hand-crafted features, such as LBP-TOP~\cite{zhao2007dynamic}, Bi-WOOF~\cite{liong2018less}, and MDMO~\cite{liu2015main}. Although these methods achieved certain success, their performance was limited by the complexity of feature design and adaptability to data variations. 
With the rise of deep learning, many researchers turned to CNNs for ME feature extraction. Liong et al.~\cite{liong2019shallow} employed a shallow three-stream CNN structure to separately process features from optical flow and optical strain in three dimensions. Zhou et al.~\cite{zhou2022feature} used a two-stream Inception network to learn expression-shared features, followed by Softmax attention for learning expression-specific features and feature fusion. Wang et al.~\cite{wang2018micro} proposed TLCNN, which combines Deep CNN for spatial features and LSTM for temporal modeling.
MERSiamC3D~\cite{zhao2021two} and ME-PLAN~\cite{zhao2022meplan} both utilized 3D-CNNs for spatiotemporal feature extraction from ME frame sequences.
Recently, ViTs have gained traction in MER. Zhang et al.~\cite{zhang2022short} completely abandoned traditional CNNs, relying solely on ViTs and LSTMs to achieve superior performance. Wang et al.~\cite{wang2024htnet} proposed HTNet, which stacks multiple hierarchical Transformer layers for feature extraction at different scales. In contrast to these approaches, our method utilizes SSMs for local feature extraction, combined with self-attention mechanisms to integrate global dependencies, enabling comprehensive modeling of fine-grained ME features across multiple receptive fields.

\begin{table*}[t]

\caption{The detailed parameters setting of each stage in MERba. Stage 1 is a FFE stage and Stage 2, 3, 4 are LGFI stages.}
\setlength{\belowdisplayskip}{-0.2cm}
\centering
\label{tab:parameters}
% \begin{threeparttable}
\begin{tabular}{c| c| c| c| c| c}
  \Xhline{3\arrayrulewidth}
  & Block & Depth & Window Size & Input Size & Output Size\\
  \hline
  Patch Embed & Conv & - & - & $224 \times 224 \times 3$ & $56 \times 56 \times 128$ \\ \hline
  Stage 1 & Conv & $N_1=3$ & - & $56 \times 56 \times 128$ & $28 \times 28 \times 256$ \\ \hline
  \multirow{2}*{Stage 2} & MERba Local Extractor & $N_2=2$ & $7 \times 7$ &\multirow{2}*{$28 \times 28 \times 256$} & \multirow{2}*{$14 \times 14 \times 512$} \\ \cline{2-4}
   & Global Self-attention & 2 & $28 \times 28$ & & \\
  \hline
  \multirow{2}*{Stage 3} & MERba Local Extractor & $N_3=6$ & $7 \times 7$ & \multirow{2}*{$14 \times 14 \times 512$} & \multirow{2}*{$7 \times 7 \times 1024$} \\ \cline{2-4}
   & Global Self-attention & 2 & $14 \times 14$ & & \\
   \hline
  \multirow{2}*{Stage 4} & MERba Local Extractor & $N_4=4$ & $7 \times 7$ & \multirow{2}*{$7 \times 7 \times 1024$} & \multirow{2}*{$1 \times 1 \times 1024$} \\ \cline{2-4}
   & Global Self-attention & 2 & $7 \times 7$ & & \\\hline
   % Classifier & Linear & - & - & $1024$ (flattened) & $C$ \tnote{1}\\
  \Xhline{3\arrayrulewidth}
\end{tabular}
% \begin{tablenotes}
% \footnotesize
% \item[1] $C$ represents the number of classes.
% \end{tablenotes}
% \end{threeparttable}
% \vspace{-0.5cm}
\end{table*}

\subsection{Scanning in Mamba}
To enhance the spatial perception of Mamba, many researchers have focused on the design of scanning methods. ViM~\cite{zhu2024vision} treated the image as 2D patch sequence and applied bi-directional scanning to capture global contextual information. VMamba~\cite{liu2024vmamba} scanned bi-directionally along both horizontal and vertical axes, further enriching spatial information extraction. %RS-Mamba~\cite{2024} introduces Omni-directional Selective Scanning, adding diagonal and anti-diagonal scanning directions to the VMamba framework, thereby more comprehensively covering image features.
LocalMamba~\cite{huang2025localmamba} not only performed bi-directional scanning globally but also within small local windows, and customized the scanning direction for each layer. PlainMamba~\cite{yang2024plainmamba} used continuous 2D scanning, maintaining the spatial continuity of adjacent tokens when transitioning between rows or columns. In contrast to above mentioned representative scanning methods, the asymmetric multi-scanning strategy proposed in this paper fully considers the inherent properties of the human face and effectively models spatial relationships for MER tasks while avoiding redundancy.

\section{Methodology}
%In Section~\ref{sec:pre}, we briefly review the core aspects of SSM and MambaVision. Our MERba model builds upon MambaVision, with critical redesigns tailored to the unique characteristics of MEs. As shown in Figure~\ref{fig:pipeline}, the overall framework of MERba is presented. We use TV-L1~\cite{zach2007duality} optical flow extracted from the onset and apex frames as input features $\textbf{x}^{OF}=(u,v,m)^T \in \mathbb{R}^{H\times W\times 3}$, where $u$ and $v$ represent the horizontal and vertical components, respectively, and $m=\sqrt{u^2+v^2}$. After an initial feature extraction through a stem and the convolutional layers of Stage 1, which are consistent with MambaVision-B, the data flows through three LGFI stages with different receptive fields (Section~\ref{sec:LGFI}). The local mixer extractors within these LGFI stages incorporate an asymmetric multi-scanning strategy (Section~\ref{sec:scan}) to enhance local spatial awareness. A downsampling operation is applied between each two stages, reducing the spatial resolution by half while doubling the number of channels. At the end of Stage 4, we employ 2D batchnorm and average pooling instead of downsampling. Finally, a linear classifier is used for classification, and debiased inference (Section~\ref{sec:infer}) is applied to alleviate overfitting to the training set distribution.
%In Section~\ref{sec:pre}, we briefly review the core principles of SSM and MambaVision, upon which our MERba model is built with key redesigns to address the unique characteristics of MEs.
The overall framework of MERba, as illustrated in Figure~\ref{fig:pipeline}, primarily consists of one Fast Feature Extraction (FFE) Stage and three Local-Global Feature Integration (LGFI) Stages.
% The overall framework of MERba is illustrated in Figure~\ref{fig:pipeline}. 
Specifically, we use TV-L1~\cite{zach2007duality} optical flow extracted from the onset and apex frames of ME sequences as input, denoted as $\textbf{x}^{OF}=(u,v,m)^T \in \mathbb{R}^{H\times W\times 3}$, where $u$ and $v$ represent the horizontal and vertical components, respectively, and $m=\sqrt{u^2+v^2}$.
$\textbf{x}^{OF}$ is first tokenized into non-overlapping patches and projected into a higher-dimensional embedding space via a convolutional patch embedding layer. 
After patch embedding, initial feature extraction is performed through $N_1$ CNN layers of FFE Stage 1, consistent with MambaVision-B~\cite{hatamizadeh2025mambavision}. Then features progress through LGFI Stages 2, 3, and 4, each designed with different receptive fields (Section~\ref{sec:LGFI}). Within these stages, MERba local extractors employ an asymmetric multi-scanning strategy (Section~\ref{sec:scan}) to enhance local spatial perception. Finally, when fine-grained recognition is required, classification is performed via the Dual-Granularity Classification Module (Section~\ref{sec:DGCM}), comprising parallel linear heads for coarse and fine emotional granularity. %, with debiased inference (Section~\ref{sec:infer}) applied to mitigate overfitting.

\subsection{Multi-Receptive Local-Global Feature Integration Stages} \label{sec:LGFI}

To effectively extract local motion features at different scales from ME samples, while simultaneously modeling global dependencies among different local regions within each scale, we propose the LGFI stages. These stages are applied as the second to fourth stages of the proposed MERba model.

\subsubsection{Multi-Receptive Field Feature Extraction}
Differing from the Mixer and Self-Attention hybrid architecture used in the original MambaVision, at each LGFI stage $k$, the input feature map $\textbf{x}_k^{in} \in \mathbb{R}^{H_k \times W_k \times D_k}$ is divided into $S_k$ non-overlapping local windows $\textbf{w}_k^{(i)} \in \mathbb{R}^{w_H \times w_W \times D_k}$, where $i \in \{1, 2, \dots, S_k\}$. The number of windows is calculated as:
\begin{equation} 
\setlength{\abovedisplayskip}{6pt}
\setlength{\belowdisplayskip}{6pt}
S_k = \frac{H_k \cdot W_k}{w_H \cdot w_W}, 
\end{equation}
where $H_k$ and $W_k$ represent the spatial height and width of the feature map at stage $k$, while $D_k$ is the channel dimension. Notably, the window size is fixed across all LGFI stages, with $w_H = w_W = 7$.

Through the downsampling operations between stages, the spatial resolution of the feature map is halved at each subsequent stage, effectively increasing the receptive field of each local window. By the fourth stage, the local window size encompasses the entire feature map, allowing global feature extraction. This progression enables a transition from fine-grained local motion feature learning to a comprehensive global understanding of facial features.
Table~\ref{tab:parameters} provides the detailed parameter configurations for each stage in MERba.

\subsubsection{Localized ME Movement Learning}

Within each local window $\textbf{w}_k^{(i)}$ at stage $k$, fine-grained motion features are extracted independently using $N_k$ MERba local extractor blocks, each followed by MLP blocks. The extraction process leverages our asymmetric multi-scanning strategy, which applies four distinct scanning directions in parallel to capture diverse spatial contiguity relationships.

First, the spatial arrangement of each local window $\textbf{w}_k^{(i)}$ is rearranged into 1D sequences $\textbf{s}_k^{(i,s)} \in \mathbb{R}^{T_w \times D_k}$, where $T_w = w_H \cdot w_W$ and $s \in \{a, b, c, d\}$ represents each scanning direction. Each sequence is then processed by the Mixer block to extract local spatial features. The Mixer we used is the same as MambaVision, but we have moved the last Linear layer to the end of our local extractor, as shown in Figure \ref{fig:pipeline}. The Mixer operation can be formulated as: 
%\vspace{-0.2cm}
\begin{equation} 
\setlength{\abovedisplayskip}{6pt}
\setlength{\belowdisplayskip}{6pt}
\textbf{z}_k^{(i,s)} = \text{MixerBlock}(\textbf{s}_k^{(i,s)}),
\end{equation}
where $\textbf{z}_k^{(i,s)} \in \mathbb{R}^{T_w \times D_k}$ denotes the intermediate representation of the sequence.
Subsequently, the sequences are reversed to their original spatial arrangement, reconstructing the corresponding 2D feature maps $\textbf{f}_k^{(i,s)} \in \mathbb{R}^{w_H \times w_W \times D_k}$. The four reconstructed maps are then fused together as: 
\begin{equation} 
\setlength{\abovedisplayskip}{6pt}
\setlength{\belowdisplayskip}{6pt}
\textbf{f}_k^{(i)} = \sum_{s\in\{a,b,c,d\}} \textbf{f}_k^{(i,s)}.
\end{equation}
The fused feature map $\textbf{f}_k^{(i)}$ is further refined by a Linear layer and a MLP block, producing the final localized feature representation for each local window $\textbf{y}_k^{(i)} \in \mathbb{R}^{w_H \times w_W \times D_k}$: 
\begin{equation} 
\setlength{\abovedisplayskip}{6pt}
\setlength{\belowdisplayskip}{6pt}
\textbf{y}_k^{(i)} = \text{MLP}(\text{Linear}(\textbf{f}_k^{(i)})).
\end{equation} 

\subsubsection{Global Dependency Modeling}

Once all local windows are processed, their features  ${\textbf{y}_k^{(i)}}, i\in\{1, 2, \dots, S_k\}$, are concatenated and spatially restructured back into the original order to form a complete feature map $\textbf{x}_k^{local} \in \mathbb{R}^{H_k \times W_k \times D_k}$. 
%\begin{equation} \textbf{x}_k^{local} = \text{Restructure}({\textbf{y}_k^{(1)}, \textbf{y}_k^{(2)}, \dots, \textbf{y}_k^{(S_k)}}). \end{equation}

To capture global dependencies across the local windows, two blocks each comprising a Multi-Head Self-Attention (MHSA) layer followed by a MLP are applied to $\textbf{x}_k^{local}$, resulting in the global ME feature map $\textbf{x}_k^{global} \in \mathbb{R}^{H_k \times W_k \times D_k}$: 
\begin{equation} 
\setlength{\abovedisplayskip}{6pt}
\setlength{\belowdisplayskip}{6pt}
\textbf{x}_k^{global} = \text{MLP}(\text{MHSA}(\textbf{x}_k^{local})).
\end{equation} 
Next, $\textbf{x}_k^{global}$ undergoes a downsampling operation to reduce spatial resolution by half while doubling channel dimensions, yielding the final output feature map $\textbf{x}_k^{out} \in \mathbb{R}^{\frac{H_k}{2} \times \frac{W_k}{2} \times 2 D_k}$, which also serves as the input $\textbf{x}_{k+1}^{in}$ of the next stage $k+1$. Notably, for Stage 4, we apply 2D batch normalization and average pooling instead of further downsampling, resulting in $\textbf{x}_{4}^{out}\in \mathbb{R}^{1 \times 1 \times D_4}$, where $D_4=D_3$.
%\begin{equation} \textbf{x}_k^{out} = \text{DownSample}(\textbf{x}_k^{global}), \end{equation}
%where $\textbf{x}_k^{out} $ also serves as the input of the next stage $\textbf{x}_{k+1}^{in}$.

\begin{figure}[t]
%\vspace{-0.2cm}
% \setlength{\belowcaptionskip}{-0.5cm} %缩小caption和下方文字的距离
    \centering
    \includegraphics[width=0.48\textwidth]{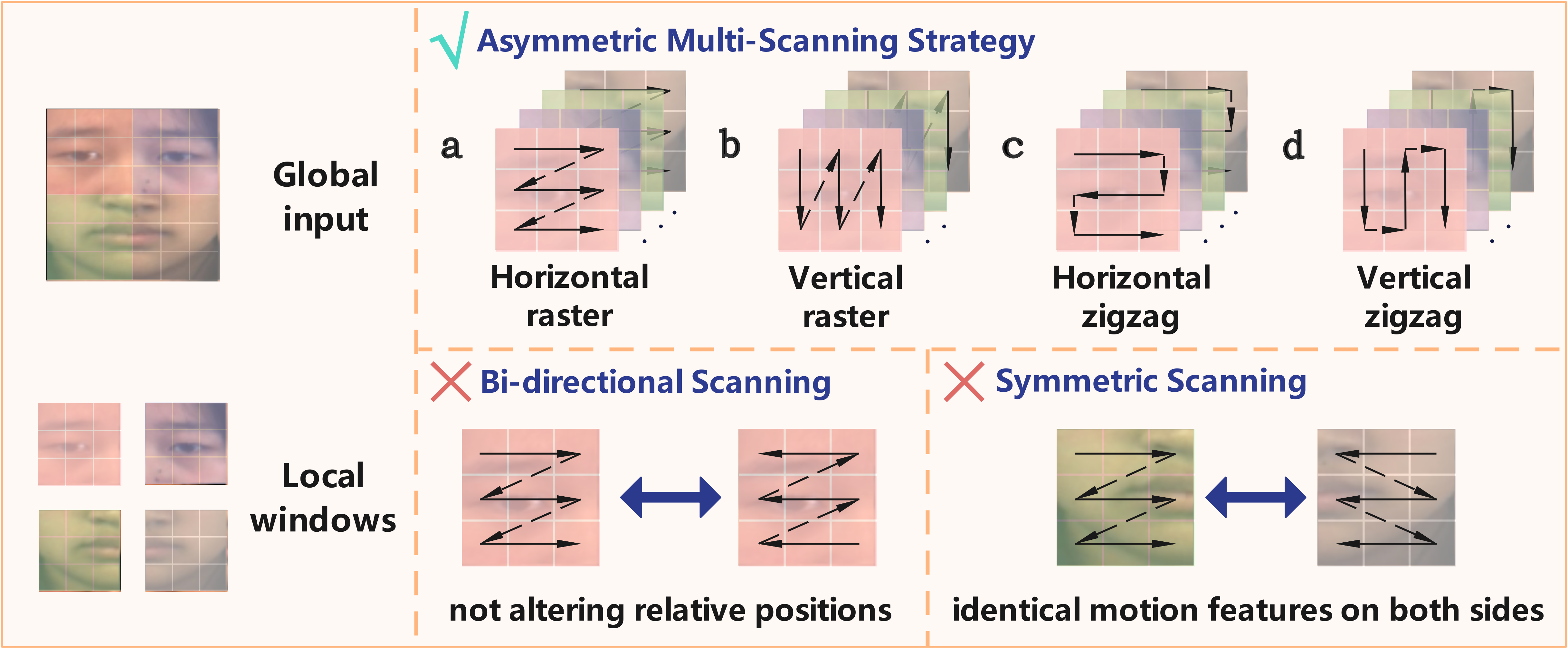}
    % \vspace{-0.1cm}
    \caption{Illustration of Asymmetric Multi-Scanning Strategy. The four scanning directions chosen in this strategy do not exhibit any bi-directional or symmetric relationships, effectively avoiding redundant scanning for human faces.}
    \label{fig:scan}
% \vspace{-0.5cm}
\end{figure}

\subsection{Asymmetric Multi-Scanning Strategy} \label{sec:scan}
%Our Asymmetric Multi-Scanning Strategy is designed to efficiently capture diverse spatial dependencies while avoiding redundancy. 
Existing scanning strategies, when involving bi-directional or symmetric directions, often lead to redundancy applying to MER. Specifically, bi-directional scanning involves reversing the order of a 1D sequence, but the relative positional relationships between tokens remain unchanged. Symmetric scanning refers to two different scanning orders derived from the left-right symmetry of a 2D image. However, when scanning within a local window, the motion features obtained from the symmetrical left and right facial regions are identical.

Therefore, as shown in Figure~\ref{fig:scan}, we employ four distinct scanning directions including \textbf{horizontal raster} ($a$), \textbf{vertical raster} ($b$), \textbf{horizontal zigzag} ($c$), and \textbf{vertical zigzag} ($d$) to process each local window $\textbf{w}_k^{(i)}$. Each scanning direction operates independently, starting at the top-left token of the window and proceeding according to its respective pattern:
\begin{itemize}
\item \textbf{Raster Scanning}: The sequence moves row-by-row (horizontally) or column-by-column (vertically), resetting to the start of the next row or column when switching.

\item \textbf{Zigzag Scanning}: The sequence alternates directions within rows (horizontally) or columns (vertically), ensuring that adjacent tokens remain spatially close even during direction changes.
\end{itemize}

By combining these four complementary scanning directions, the strategy captures diverse spatial relationships and effectively models the subtle, localized motion features critical for MER. Importantly, this approach avoids the redundancy introduced by bi-directional or symmetric scanning, as each direction focuses on unique spatial dependencies.

\subsection{Dual-Granularity Classification Module} 
\label{sec:DGCM}
We introduce a Dual-Granularity Classification Module (DGCM) to handle high inter-class similarity in fine-grained MER scenarios. Specifically, unlike conventional approaches that rely on a single linear classification head with an output dimension equal to the total number of emotion classes, DGCM adopts a coarse-to-fine classification paradigm via two parallel linear classifiers: a coarse-grained head $h_{\text{coarse}}$ and a fine-grained head $h_{\text{fine}}$. $h_{\text{coarse}}$ first categorizes each ME sample into one of several broad affective groups (e.g., \textit{negative}, \textit{positive}, \textit{surprise}, and optionally \textit{others} or \textit{contempt}). When a sample is predicted as \textbf{negative}, $h_{\text{fine}}$ is activated to further disambiguate among multiple visually similar negative emotions, such as \textbf{anger}, \textbf{disgust}, \textbf{fear}, and \textbf{sadness}. This coarse-to-fine mechanism encourages the model to focus its representational capacity where fine distinctions are most needed, while maintaining robustness and generalizability.

To effectively optimize this hierarchical prediction process, we introduce a two-branch training objective with progressive weighting between the coarse and fine levels. Specifically, during training, the outputs $\hat{y}_{\text{coarse}}$ and $\hat{y}_{\text{fine}}$ from both classification heads are used to compute two cross-entropy losses: $\mathcal{L}_{\text{coarse}}$ and $\mathcal{L}_{\text{fine}}$, respectively. The total training loss $\mathcal{L}_{\text{cls}}$ is progressively weighted to balance learning dynamics:
\begin{equation} 
\setlength{\abovedisplayskip}{6pt}
\setlength{\belowdisplayskip}{6pt}
\mathcal{L}_{\text{cls}} = 0.5 \cdot (\mathcal{L}_{\text{coarse}}+\alpha\cdot\mathcal{L}_{\text{fine}}),
\end{equation} 
where the weighting factor $\alpha$ increases with training:
\begin{equation} 
\setlength{\abovedisplayskip}{6pt}
\setlength{\belowdisplayskip}{6pt}
\alpha = \min (0.5+\frac{2.0\cdot\text{epoch}}{\text{total\_epochs}},\ 2.0).
\end{equation} 
Note that $\mathcal{L}_{\text{fine}}$ is computed only when the ground-truth label belongs to the negative emotion category.

At inference time, the final label $\hat{y}$ is determined via:
\begin{itemize}
\item If the coarse prediction $\hat{y}_{\text{coarse}}$ isn't \textbf{negative}, then $\hat{y}$ is directly mapped from the coarse class to the original 7-class label space.

\item Otherwise, the final label is derived from the fine prediction $\hat{y}_{\text{fine}}$ through a predefined fine-to-full label mapping.
\end{itemize}

\begin{table}[t]
%\small
\centering
%\captionsetup{font={small}}
\setlength{\abovecaptionskip}{0.1cm} %缩小caption和表格之间的距离
\setlength{\belowcaptionskip}{-0.1cm} %缩小caption和上方文字的距离
\begin{threeparttable}
\caption{Sample distribution of DFME-public and MMEW datasets.\tnote{1} }
\label{tab:DFME}
\begin{tabular}{w{c}{0.6cm}|w{c}{0.7cm}|w{c}{0.3cm} w{c}{0.3cm} w{c}{0.3cm} w{c}{0.3cm} w{c}{0.3cm} w{c}{0.3cm} w{c}{0.3cm} |w{c}{0.4cm}}
  \Xhline{3\arrayrulewidth}
  \multicolumn{2}{c|}{ } & \textit{Ang} & \textit{Con} & \textit{Dis} & \textit{Fea} & \textit{Hap} & \textit{Sad} & \textit{Sur} & Total \\
  \hline
  \multirow{3}*{DFME} & Train & 161 & 100 & 548 & 265 & 206 & 278 & 298 & 1856 \\ 
   & Test A & 39 & 34 & 129 & 62 & 63 & 46 & 101 & 474\\
   & Test B & 41 & 37 & 58 & 38 & 42 & 35 & 48 & 299\\
  \hline
  \multicolumn{2}{c|}{MMEW} & 8 & - & 72 & 16 & 36 & 13 & 89 & 234 \\
  \Xhline{3\arrayrulewidth}
\end{tabular}

% \vspace{-0.1cm}
\begin{tablenotes}
\footnotesize
\item[1] \textit{Ang}-Anger, \textit{Con}-Contempt, \textit{Dis}-Disgust, \textit{Fea}-Fear, \textit{Hap}-Happiness, \\ \textit{Sad}-Sadness, \textit{Sur}-Suprise.
\end{tablenotes}
\end{threeparttable}
% \vspace{-0.2cm}     % reduce space between the table and the next section
\end{table}

\begin{table}[t]
%\small
%\captionsetup{font={small}}
\setlength{\abovecaptionskip}{0.1cm} %缩小caption和表格之间的距离
\setlength{\belowcaptionskip}{-0.1cm} %缩小caption和上方文字的距离
\begin{center}
\caption{Sample distribution of 3DB-Combined dataset.}
\label{tab:3DB}
\begin{tabular}{c| c c c| c}
  \Xhline{3\arrayrulewidth}
   & SMIC-HS & CASME II & SAMM & Combined
  \\
  \hline
  \textit{Negative} & 70 & 88 & 92 &250 \\
  
  \textit{Positive} & 51 & 32 & 26 &109 \\
  
  \textit{Surprise} & 43 &25 & 15 &83 \\
  \hline
  Total    & 164 &145 &133 &442\\
  \Xhline{3\arrayrulewidth}
\end{tabular}
\end{center}
% \vspace{-0.2cm}     % reduce space between the table and the next section
\end{table}

\begin{table*}[t]
\setlength{\abovecaptionskip}{0.1cm} %缩小caption和表格之间的距离
\setlength{\belowcaptionskip}{-0.6cm} %缩小caption和上方文字的距离

%\captionsetup{font={small}}
\begin{center}
\caption{Experimental results compared with SOTA methods on 3DB-Combined datasets.}

\label{tab:result}
\begin{threeparttable}
\begin{tabular}{c| c| c c c c c c c c}
  \Xhline{3\arrayrulewidth}
  % \toprule[1.5pt]
  \multirow{2}*{MER Methods} & \multirow{2}*{Backbone} & \multicolumn{2}{c}{Combined} & \multicolumn{2}{c}{SMIC-HS}& \multicolumn{2}{c}{CASME II} & \multicolumn{2}{c}{SAMM} 
  \\
  \cline{3-10}
  % \cmidrule(r){3-10}
  & & UF1 & UAR & UF1 & UAR & UF1 & UAR & UF1 & UAR \\
  \Xhline{3\arrayrulewidth}
  LBP-TOP (2014) \cite{zhao2007dynamic}& Hand-crafted & 0.5882 & 0.5785& 0.2000 & 0.5280 & 0.7026 & 0.7429 & 0.3954 &0.4102\\

  Bi-WOOF (2018) \cite{liong2018less} & Hand-crafted &0.6296 &0.6227 & 0.5727 & 0.5829 & 0.7805& 0.8026 & 0.5211 & 0.5139\\
  \hline
  STSTNet (2019) \cite{liong2019shallow} & 2D-CNN &0.7353&0.7605& 0.6801 & 0.7013 & 0.8382 & 0.8686&  0.6588 &0.6810\\
  
  FeatRef (2022) \cite{zhou2022feature} & 2D-CNN &0.7838&0.7832& 0.7011 & 0.7083 & 0.8915 & 0.8873  &0.7372 &0.7155\\
  
  ME-PLAN (2022) \cite{zhao2022meplan}& 3D-CNN &0.7715&0.7864 & 0.7127 & 0.7256 & 0.8632 & 0.8778 &  0.7164 & 0.7418\\
  \hline
  
  SLSTT (2022) \cite{zhang2022short} & ViT+LSTM &0.8160&0.7900 & 0.7400 & 0.7200 & 0.9010 & 0.8850 &  0.7150 & 0.6430\\

  MFDAN (2024) \cite{cai2024mfdan} & ViT & 0.8453 &\underline{0.8688} & 0.6815 & 0.7043 & 0.9134 & 0.9326 & 0.7871 & \underline{0.8196}\\

  HTNet (2024) \cite{wang2024htnet} & ViT &\underline{0.8603} &0.8475 & \underline{0.8049} & \underline{0.7905} & \textbf{0.9532} & \textbf{0.9516} &  \underline{0.8131} & 0.8124\\  
  \hline

   MambaVision-B\tnote{1} ~(2024) \cite{hatamizadeh2025mambavision} & SSM+ViT & 0.8193  & 0.8227  & 0.7382  & 0.7435  & 0.9268 & 0.9327 & 0.7617  & 0.7520  \\
  
  \textbf{MERba (ours)} & SSM+ViT & \textbf{0.8840} & \textbf{0.8777} & \textbf{0.8312} & \textbf{0.8309} & \underline{0.9498} & \underline{0.9470} & \textbf{0.8701} & \textbf{0.8423} \\

  \Xhline{3\arrayrulewidth}
  %\bottomrule[1.5pt]
\end{tabular}

\begin{tablenotes}
\footnotesize
\item[1] MambaVision-B is reproduced with the same parameters as MERba.
\end{tablenotes}

\end{threeparttable}
\end{center}
% \vspace{-0.4cm}     % reduce space between the table and the next section
\end{table*}

\section{Experiments}
In this section, we first provide an overview of the datasets and evaluation metrics used in our experiments. We then detail the implementation settings and present the results of our method. Finally, we conduct an ablation study to analyze the contributions of each component in our model, and include a case study to provide visual explanations for how the model captures key emotional cues.

\subsection{Datasets and Evaluation Metrics}
%介绍3DB和DFME两个，包括两种不同的验证方式
To ensure a comprehensive and fair comparison, we conducted experiments on three representative MER datasets, 3DB-Combined (3 classes), DFME-public (7 classes), and MMEW (6 classes). Detailed sample distributions of the three datasets involved in our experiments are provided in Table \ref{tab:DFME} and \ref{tab:3DB}.
\begin{itemize}
\item \textbf{3DB-Combined} is a composite dataset of SMIC-HS~\cite{li2013spontaneous}, CASME II~\cite{yan2014casme}, and SAMM~\cite{davison2016samm}. We employed Leave-One-Subject-Out~(LOSO) cross-validation with Unweighted F1-score (UF1) and Unweighted Average Recall (UAR) as evaluation metrics in line with MEGC2019 protocols~\cite{see2019megc}.
\item \textbf{DFME-public} refers to the publicly available portion of the DFME dataset~\cite{zhao2023dfme}, subject-independently divided into a fixed training set and two test sets. The model was trained using the entire training set and inference was conducted separately on Test A and Test B. Following \cite{zhao2024dynamic}, we used UF1, UAR, and Accuracy (ACC) as evaluation metrics.
\item \textbf{MMEW}~\cite{ben2021video} contains 300 ME samples across 7 emotion categories. To align with the experimental protocol in the original paper, we excluded the \textit{others} category. The remaining 234 samples, covering 6 emotion classes, were used for subject-independent 5-fold cross-validation. ACC was adopted as the evaluation metric, consistent with the benchmark setting.
\end{itemize}

\subsection{Implementation Settings}
%提3DB初始化权重是来自DFME
For all datasets, we first apply face detection, alignment, and cropping, following the same procedure as~\cite{zhao2023dfme}. Subsequently, all images are resized to a uniform resolution of $224\times224$. For the SMIC-HS dataset, which lacks apex frame labels, we employ the apex frame detection method from~\cite{zhao2022meplan} to generate the labels. During training, the only data augmentation applied is random horizontal flipping. The optimizer used is AdamW, with the learning rate warmed up over the first 5 epochs to $5e-4$, followed by a cosine decay scheduler, with a cooldown period over the last 10 epochs. The weight decay is set to 0.05, and the classification loss function is cross-entropy loss. The total training epochs are 200 for DFME-public and each fold of MMEW. For 3DB-combined, the number of epochs per subject is 70. Early stopping is employed during training to prevent overfitting, and the dropout rate is set to 0.1. All experiments are conducted using the PyTorch framework, with training and inference performed on a single NVIDIA A100 Tensor Core GPU, and the training batch size is set to 128.

\subsection{Experimental Results}

\subsubsection{3DB-Combined}
The results are compared across a range of baseline methods, spanning hand-crafted~\cite{zhao2007dynamic,liong2018less}, 2D-CNN~\cite{liong2019shallow,zhou2022feature}, 3D-CNN~\cite{zhao2022meplan}, and Transformer-based~\cite{zhang2022short,cai2024mfdan,wang2024htnet} approaches. MambaVision-B~\cite{hatamizadeh2025mambavision} and our MERba represent the latest models using SSM+ViT backbones. Since the 3DB-Combined dataset adopts a 3-class protocol without fine-grained negative emotion labels, we report results using MERba without the DGCM module. As shown in Table~\ref{tab:result}, our method achieves SOTA results on SMIC-HS, SAMM, and the overall 3DB-combined dataset. On CASME II, MERba performs competitively with HTNet, trailing by less than 0.5\% in both UF1 and UAR. Notably, on the SAMM dataset, which has a more diverse ethnic groups, our model surpasses HTNet by a large margin of 5.7\% UF1, demonstrating stronger generalization across demographic variation. Moreover, MERba achieves this performance with only 101.21M parameters—over four times smaller than HTNet’s 438.51M—highlighting its superior parameter efficiency.

\begin{table}[t]
\caption{Comparison with SOTA methods on DFME-public.}
\centering
\label{tab:result_dfme}
\begin{threeparttable}
\begin{tabular}{c| c| c c c}
  \Xhline{3\arrayrulewidth}
  MER Methods & Test Set &  UF1 & UAR & ACC\\
  
  \hline
  FeatRef (2022)\cite{zhou2022feature} & \multirow{7}*{Test A} & 0.3410 &0.3686 & \underline{0.5084}\\
  Wang et al. (2024)\cite{zhao2024dynamic} &  & 0.4067 & 0.4074 & 0.4641\\
  He et al. (2024)\cite{zhao2024dynamic} &  & \underline{0.4123} & \underline{0.4210} & 0.4873 \\
  HTNet\tnote{1} ~(2024)\cite{wang2024htnet} & & 0.3736 & 0.3821 & 0.4768\\
  MambaVision-B\tnote{1} ~(2024)~\cite{hatamizadeh2025mambavision} & & 0.4002 & 0.4064 & 0.4578\\
  \textbf{MERba-base\tnote{2}~ (ours)} & & 0.4101 & 0.4123 & 0.4831 \\
  \textbf{MERba-DGCM (ours)} & & \textbf{0.4332} & \textbf{0.4287} & \textbf{0.5232} \\  
  
  \hline
  FeatRef (2022)\cite{zhou2022feature} & \multirow{7}*{Test B} & 0.2875 & 0.3228 & 0.3645\\
  Wang et al. (2024)\cite{zhao2024dynamic} & & 0.3534 & 0.3661 & 0.3813\\
  He et al. (2024)\cite{zhao2024dynamic} & & 0.4016 & 0.4008 & 0.4147 \\
  HTNet\tnote{1} ~(2024)\cite{wang2024htnet} & & 0.4076 & 0.4062 & 0.4214\\
  MambaVision-B\tnote{1} ~(2024)~\cite{hatamizadeh2025mambavision} & & 0.3929 & 0.3858 & 0.4080\\
  \textbf{MERba-base\tnote{2}~ (ours)} & & \underline{0.4114} & \underline{0.4219} & \textbf{0.4415} \\
  \textbf{MERba-DGCM (ours)} & & \textbf{0.4302} & \textbf{0.4263} & \textbf{0.4415} \\  
  
  \Xhline{3\arrayrulewidth}
\end{tabular}
\begin{tablenotes}
\footnotesize
\item[1] HTNet is reproduced using the hyperparameters reported in its original paper, and MambaVision-B is reproduced with the same hyperparameters as MERba-base.
\item[2] MERba-base denotes replacing DGCM with a single linear head which predicts all seven emotion categories.
\end{tablenotes}
\end{threeparttable}
% \vspace{-0.5cm}
\end{table}

\subsubsection{DFME-public}
DFME-public test sets take FeatRef~\cite{zhou2022feature} as the baseline, and we further include the results of top two teams from the DFME Challenge at CCAC 2024, which are publicly available on the official leaderboard~\cite{zhao2024dynamic}. We have also reproduced HTNet~\cite{wang2024htnet} and MambaVision-B~\cite{hatamizadeh2025mambavision} for a more comprehensive comparison. As shown in Table~\ref{tab:result_dfme}, on both Test Set A and B, MERba with DGCM outperforms all SOTA methods, demonstrating superior performance in the challenging fine-grained 7-class MER task.

\begin{table}[t]
\caption{Comparison with SOTA methods on MMEW.}
\centering
\label{tab:result_mmew}
\begin{threeparttable}
\begin{tabular}{c| c| c}
  \Xhline{3\arrayrulewidth}
  MER Methods & Backbone &  ACC(\%)\\
  \hline
  LBP-TOP (2014)\cite{zhao2007dynamic} & Hand-crafted & 38.9 \\
  MDMO (2015)\cite{liu2015main} & Hand-crafted & 65.9 \\
  \hline
  TLCNN (2018)\cite{wang2018micro} & CNN+LSTM & 69.4 \\
  LD-FMERN (2023)\cite{ni2023diverse} & 2D-CNN & 71.7 \\
  HTNet\tnote{1} ~(2024)\cite{wang2024htnet} & ViT & 71.8 \\
  MA2MI (2024)\cite{li2024macro} & transfer learning & \textbf{75.2} \\
  \hline
  MambaVision-B\tnote{1} ~(2024)~\cite{hatamizadeh2025mambavision} & SSM+ViT & 70.1\\
  \textbf{MERba-base\tnote{2}~ (ours)} & SSM+ViT & 72.6 \\
  \textbf{MERba-DGCM (ours)} & SSM+ViT & \textbf{75.2} \\  
  
  \Xhline{3\arrayrulewidth}
\end{tabular}
\begin{tablenotes}
\footnotesize
\item[1] HTNet is reproduced using the hyperparameters reported in its original paper, and MambaVision-B is reproduced with the same hyperparameters as MERba-base.
\item[2] MERba-base denotes replacing DGCM with a single linear head which predicts all six emotion categories.
\end{tablenotes}
\end{threeparttable}
% \vspace{-0.5cm}
\end{table}

\subsubsection{MMEW}
As shown in Table~\ref{tab:result_mmew}, MERba outperforms all baseline models based on hand-crafted features, CNNs, and Transformers on the MMEW dataset. MERba with DGCM achieves the highest accuracy of 75.2\%, performing on par with MA2MI~\cite{li2024macro}, which denotes a SOTA model employing transfer learning. Notably, MA2MI leverages extensive macro-expression data for pretraining, while our MERba is trained solely on ME samples, yet reaches top-level performance. This highlights how our architecture effectively adapts to the MER task and captures fine-grained emotional distinctions, even with limited supervision.

\begin{table}[t]

\setlength{\abovecaptionskip}{0.1cm} %缩小caption和表格之间的距离
\setlength{\belowcaptionskip}{-0.1cm} %缩小caption和上方文字的距离

\begin{center}
\caption{Ablation study for Local-Global Feature Integration on DFME-public Test B.}
\label{tab:L-G}
\begin{tabular}{c| c| c c c}
  \Xhline{3\arrayrulewidth}
  Extractor & Self-Attention &  UF1 & UAR & ACC \\  
  \hline
  local & local &  0.3749 & 0.3637 & 0.3746 \\
  global & global & 0.3840 & 0.3814 & 0.4013 \\
  % global & local & 0.4210 & 0.4120 & 0.4247 \\
  \textbf{local} & \textbf{global} & \textbf{0.4302} & \textbf{0.4263} & \textbf{0.4415} \\
  \Xhline{3\arrayrulewidth}
\end{tabular}

% \vspace{-0.3cm}
\end{center}
\end{table}

\begin{table}[t]

\setlength{\abovecaptionskip}{0.1cm} %缩小caption和表格之间的距离
\setlength{\belowcaptionskip}{-0.1cm} %缩小caption和上方文字的距离
\centering
\begin{threeparttable}

\caption{Ablation study for Asymmetric Multi-Scanning on DFME-public Test B.}
\label{tab:scanning}
\begin{tabular}{c| c c c}
  \Xhline{3\arrayrulewidth}
  Directions\tnote{1} & UF1 & UAR & ACC\\
  \hline
  $a$ & 0.3814 & 0.3713 & 0.3746\\
  $a+b$ & 0.3849 & 0.3810 & 0.3946\\
  $c+d$ & 0.3929 & 0.3939 & 0.4147 \\
  $a+a_{bi}+b+b_{bi}$ & 0.4102 & 0.4020 & 0.4147 \\
  $a+a_{sy}+b+b_{sy}$ & 0.4076 & 0.4051 & 0.4181 \\  
  $a+b+c+d$ & \textbf{0.4302} & \textbf{0.4263} & \textbf{0.4415} \\
  \Xhline{3\arrayrulewidth}
\end{tabular}

\begin{tablenotes}
\footnotesize
\item[1] $a, b, c, d$ are consistent with the corresponding direction in Figure \ref{fig:scan}. The subscripts "$bi$" or "$sy$" respectively represent the scanning directions exhibiting a bi-directional or horizontal symmetric relationship with the original direction $a$ or $b$.
\end{tablenotes}
\end{threeparttable}
% \vspace{-0.5cm}
\end{table}

%列表+混淆矩阵分析
\subsection{Ablation Study}
\subsubsection{Local-Global Feature Integration}

%In this section, we examine the impact of window size for the Mixer and Self-Attention modules in each LGFI Stage. 
%Our default setting combines a local extractor with global self-attention. According to Table \ref{tab:L-G}, when we adjust the self-attention blocks with the same local window size as the MERba Local Extractors in each stage, the model performance drops significantly (UF1 down by 3.61\% and UAR by 3.29\%). This decline indicates that local self-attention struggles to model interdependencies between different facial regions during ME occurrences. Conversely, when we set the Extractor containing Mixers to a global window size to match the self-attention, MER performance also decreases (UF1 down by 2.24\% and UAR by 2.38\%). This is because it lacks fine-grained spatial learning required to detect subtle facial movements  when using global Extractors. These results highlight the effectiveness of Local-Global Feature Integration for optimal performance.
We perform an ablation study to assess the effect of our proposed LGFI stages on the DFME-public Test B dataset, as shown in Table~\ref{tab:L-G}.
On the one hand, when \textit{self-attention} is applied only within the local window like the \textit{extractors}, the model performs poorly, struggling to capture global dependencies between different facial regions during ME occurrences. On the other hand, when we discard the local window (i.e., both \textit{extractors} and \textit{self-attention} are applied to the entire feature map), the model achieves a UF1 of 0.3840 and a UAR of 0.3814. However, it still lacks the fine-grained spatial learning necessary for detecting subtle facial movements.
The best performance is achieved by combining local \textit{extractors} with global \textit{self-attention}, demonstrating that associating local feature extraction with global dependency modeling provides a substantial improvement in MER performance.

\begin{figure}[t]
    \centering
    \begin{subfigure}[b]{0.45\textwidth}
        \includegraphics[width=\textwidth]{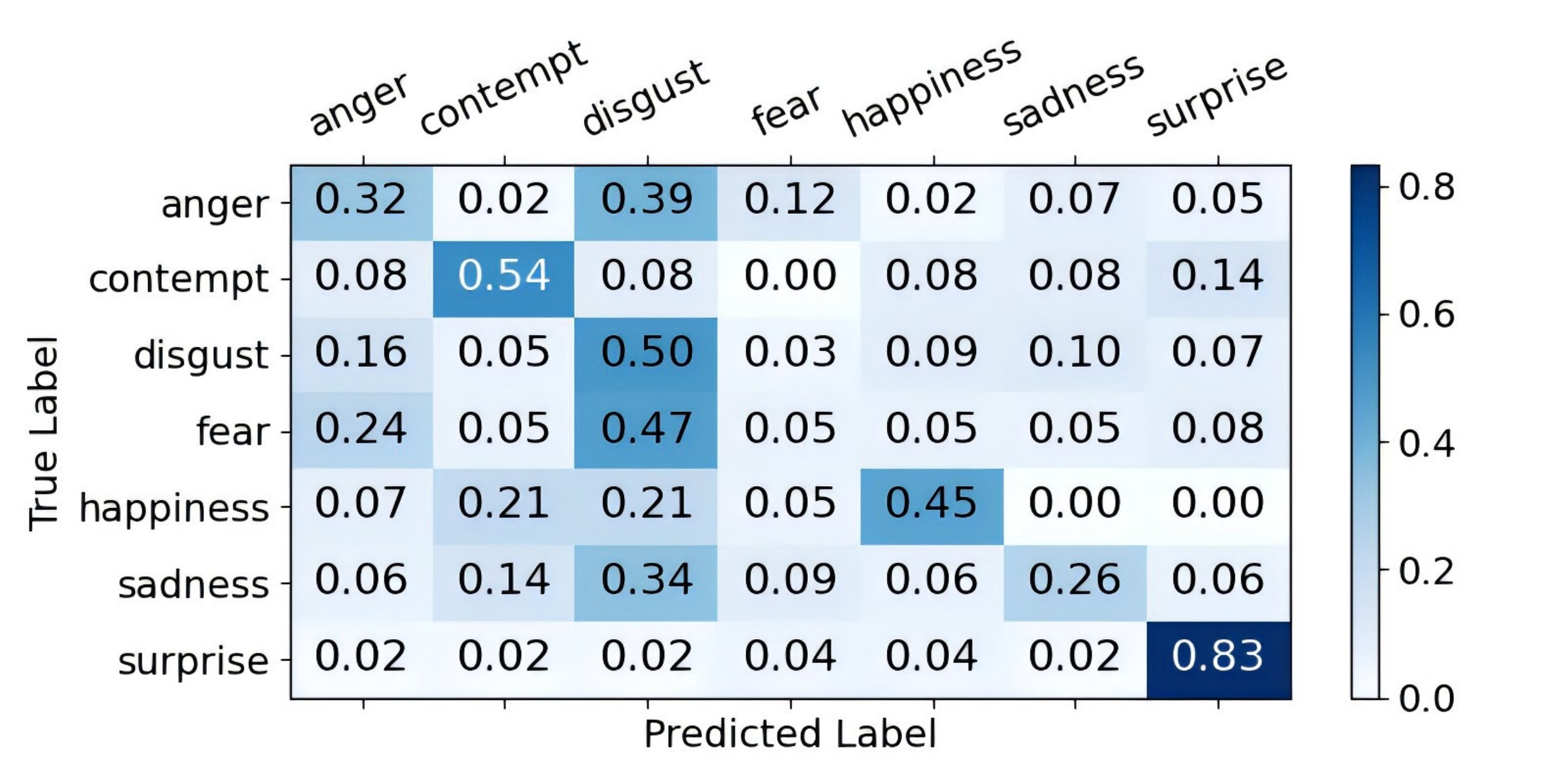}
        % \vspace{-0.9cm}
        \caption{MERba-base (w/o DGCM)}
        \label{fig:MERba-base}
    \end{subfigure}
    \\[5pt]
    \centering
    \begin{subfigure}[b]{0.45\textwidth}        \includegraphics[width=\textwidth]{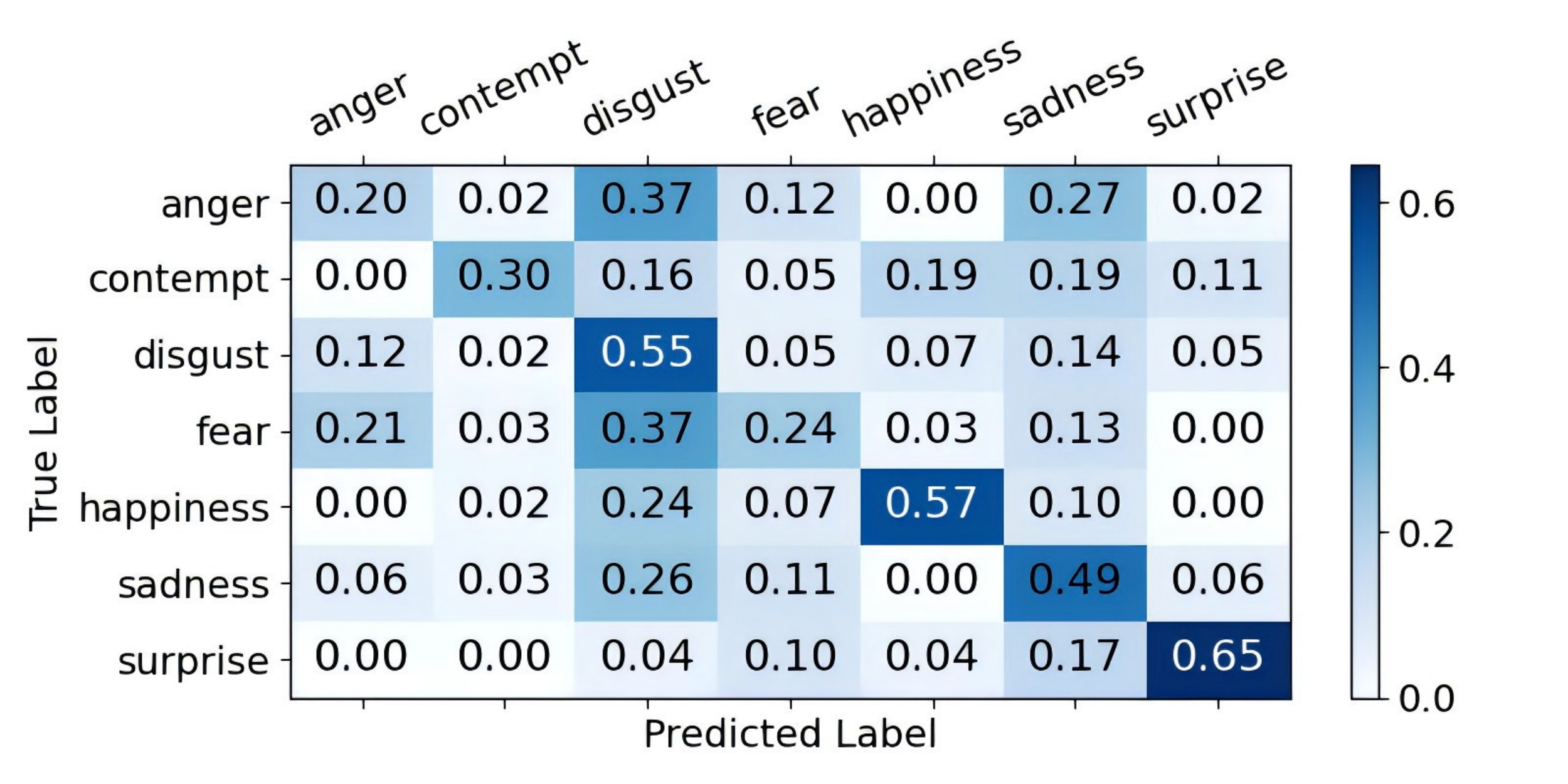}
        % \vspace{-0.9cm}
        \caption{MERba-DGCM}
        \label{fig:MERba-DGCM}
    \end{subfigure}

    \caption{Confusion Matrices on DFME-public Test B.}
    \label{fig:confusion}
% \vspace{-0.3cm}
\end{figure}

\subsubsection{Asymmetric Multi-Scanning Strategy}
We first investigate the impact of the number of scanning directions on MER performance. As shown in Table~\ref{tab:scanning}, compared with using a single ($a$) or a pair of directions ($a, b$), introducing more scanning directions leads to consistent performance gains across UF1, UAR, and ACC. This confirms that incorporating diverse scanning combinations enhances the spatial perception capability of the local extractor in ME feature extraction.
However, we also observe that adding bi-directional ($a_{bi}$, $b_{bi}$) or horizontal symmetric ($a_{sy}$, $b_{sy}$) counterparts to the base directions does not lead to significant improvements. In contrast, utilizing our proposed asymmetric multi-scanning directions ($a, b, c, d$) achieves the best performance, indicating its ability to capture the most discriminative motion patterns without introducing unnecessary directional overlap.
% We first validate the redundancy introduced by bi-directional and symmetric scanning in ME motion feature learning. Comparing the first row with the third and fourth rows of Table~\ref{tab:scanning}, the inclusion of both only slightly improves MER performance while increasing computational cost. Then we evaluate the effectiveness of different scanning direction combinations and the spatial-channel attention module.
% Specifically, compared to using only raster or zigzag scanning in two axes, combining all four scanning directions with the spatial-channel attention module achieved the best performance by capturing richer and more detailed contiguity relationships in both horizontal and vertical directions. However, when the spatial-channel attention module is removed, the model fails to effectively capture the contribution of each scanning direction, leading to a performance decrease. 
% Overall, these results demonstrate that the asymmetric multi-scanning strategy effectively minimizes the redundancy, while facilitating the model's ability to learn subtle ME motion features.

% \begin{figure}[t]
% \vspace{-0.2cm}
% \setlength{\abovecaptionskip}{0.1cm}
% \setlength{\belowcaptionskip}{-0.6cm} %缩小caption和下方文字的距离
%     \centering
%     \includegraphics[width=0.3\textwidth]{figures/plot_output.jpg}
%     \vspace{-0.1cm}
%     \caption{MER Performance with debiased inference on DFME-public Test B at different debiasing rates.}
%     \label{fig:debias}
% \end{figure}

\subsubsection{Dual-Granularity Classification Module}
In Table \ref{tab:result_dfme} and \ref{tab:result_mmew}, we report the MER performance both with and without DGCM. It can be observed that introducing DGCM consistently improves performance on both DFME-public Test A and B, yielding UF1 gains of 2.31\% and 1.88\%, respectively. 
Similarly, on the MMEW dataset, DGCM leads to a 2.6\% increase in overall accuracy, further validating its effectiveness across datasets.
To gain deeper insight into its impact, the confusion matrices in Figure~\ref{fig:confusion} visualize how DGCM alters class-wise predictions.
Notably, the number of true positives (TP) for three $negative$ emotions—$disgust$, $fear$, and $sadness$—increases significantly. In addition, the tendency to misjudge $fear$ as $anger$/$disgust$, as well as the reciprocal confusion between $anger$ and $disgust$, is noticeably alleviated. These observations confirm DGCM’s effectiveness in disambiguating fine-grained negative emotions.
However, this improvement comes with a trade-off: accuracy for certain non-negative emotions such as $surprise$ slightly decreases, possibly due to the model focusing more on subtle negative distinctions at the expense of broader category coverage.

% \subsubsection{Debiased Inference}
% We conduct a training-free debiased inference experiment on Test B of DFME-public, which has a more balanced sample distribution compared to the training set. As shown in Figure \ref{fig:debias}, debiasing significantly improves all three evaluation metrics when $\beta$ is set to 0.4. However, performance declines when $\beta$ is either too low or too high. This is because small weights do not generate sufficient disturbance, while excessively large weights lead to overcorrection. Moreover, based on the confusion matrices shown in Figure \ref{fig:TestB=0}\&\ref{fig:TestB=0.4}, we found that the \textit{Recall} for the tail classes (\textit{contempt, fear}, and \textit{sadness}) and the \textit{Precision} for the head class \textit{disgust} both increase after debiasing. This indicates that debiased inference effectively addresses the inherent issue of class imbalance in small-sample ME datasets.

\subsection{Case Study}
\begin{figure}[t]
%\vspace{-0.2cm}
% \setlength{\belowcaptionskip}{-0.5cm} %缩小caption和下方文字的距离
    \centering
    \includegraphics[width=0.44\textwidth]{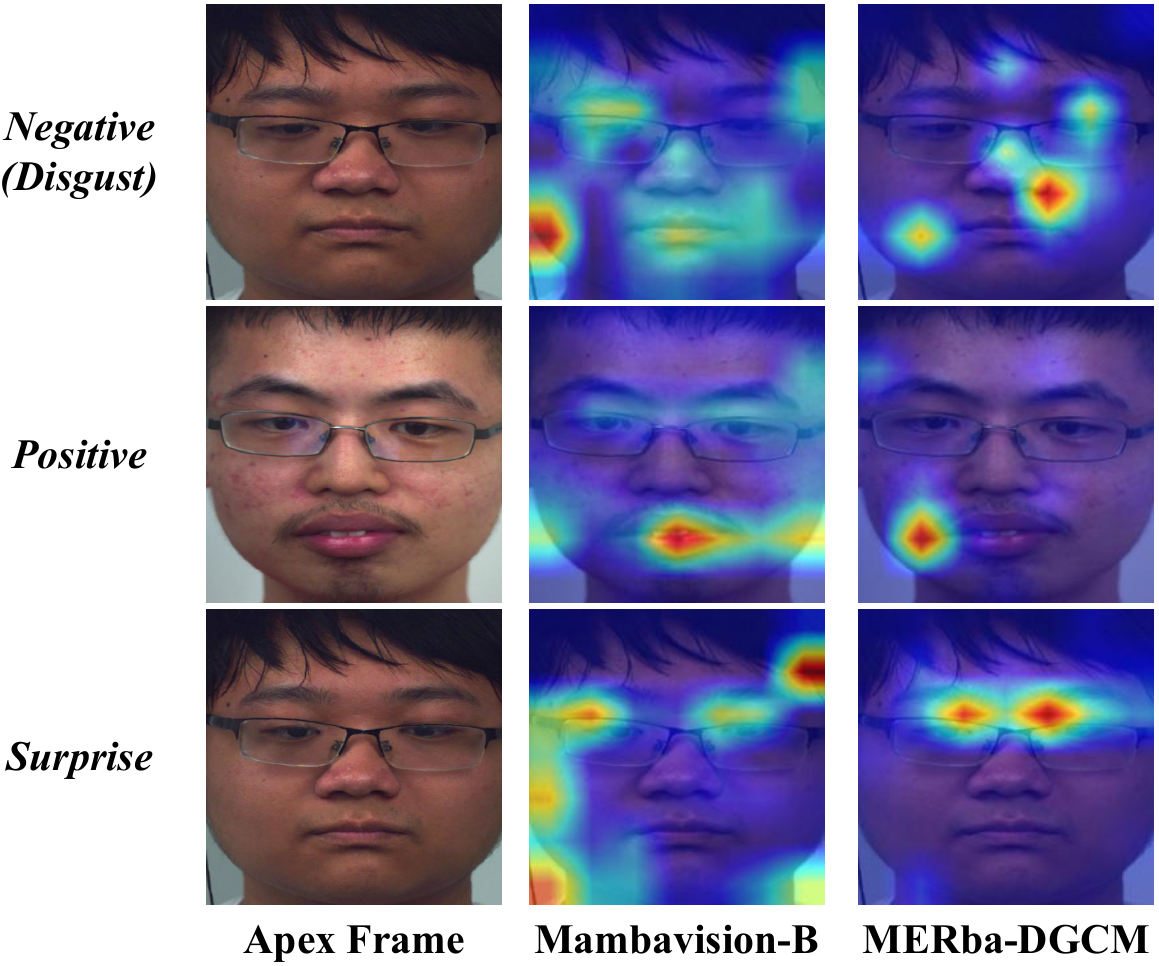}
    % \vspace{-0.1cm}
    \caption{Grad-CAM visualizations of ME samples from DFME.}
    \label{fig:gradcam}
% \vspace{-0.5cm}
\end{figure}

To better understand the inner workings of our model, we conduct a visual case study using Gradient-weighted Class Activation Mapping (Grad-CAM)\cite{selvaraju2017grad}. As shown in Figure \ref{fig:gradcam}, we compare the attention heatmaps generated by MambaVision-B and our MERba-DGCM on three unseen test samples from the DFME dataset, covering the emotion categories of \textit{negative} (\textit{disgust}), \textit{positive}, and \textit{surprise}. For both models, Grad-CAM is applied to the last layer of the feature extractor to enable a direct comparison of high-level spatial attention.

A closer look reveals that MambaVision-B tends to spread attention over irrelevant areas, sometimes mistakenly focusing on background regions, including the subject’s hair or the image boundaries. In contrast, our MERba-DGCM consistently attends to semantically meaningful facial areas, such as the nasolabial folds for \textit{disgust}, the cheek and lip corners for \textit{positive}, and the widened eye region for \textit{surprise}. These qualitative results complement our quantitative findings, confirming that MERba-DGCM captures more discriminative and interpretable motion cues through its integrated local-global modeling and coarse-to-fine classification strategy.

\section{Conclusion}
In this paper, we proposed MERba, a novel architecture for MER, designed to address key challenges in feature extraction, global dependency modeling, and high inter-class similarity of MEs. By combining local extractors with global self-attention mechanisms, MERba effectively captures both fine-grained facial features and long-range dependencies, which are crucial for recognizing MEs. The asymmetric multi-scanning strategy reduces redundancy in scanning while enhancing the model's spatial perception. 
Additionally, DGCM introduces a dual-head coarse-to-fine paradigm that disentangles emotion polarity from subtle category distinctions, enhancing recognition of highly similar MEs. 
The experimental results on benchmark MER datasets (3DB-Combined, DFME-public, and MMEW) demonstrate that MERba outperforms existing methods, setting new SOTA performance.

\bibliographystyle{IEEEbib}
\bibliography{icme2025references}

\end{document}